\documentclass{article}
\usepackage{spconf,amsmath,epsfig,multirow}
\usepackage{amssymb}
\let\OLDthebibliography\thebibliography
\renewcommand\thebibliography[1]{
  \OLDthebibliography{#1}
  \setlength{\parskip}{0pt}
  \setlength{\itemsep}{0pt plus 0.3ex}
}

\begin{document}\sloppy

\def\x{{\mathbf x}}
\def\L{{\cal L}}

\title{Deep unfolding Network for Hyperspectral Image Super-Resolution with Automatic Exposure Correction}
%
\name{Yuan Fang, Yipeng Liu, Jie Chen, Zhen Long, Ao Li, Chong-Yung Chi, Ce Zhu}
\address{}

\maketitle

\begin{abstract}
In recent years, the fusion of high spatial resolution multispectral image (HR-MSI) and low spatial resolution hyperspectral image (LR-HSI) has been recognized as an effective method for HSI super-resolution (HSI-SR). However, both HSI and MSI may be acquired under extreme conditions such as night or poorly illuminating scenarios, which may cause different exposure levels, thereby seriously downgrading the yielded HSISR. In contrast to most existing methods based on respective low-light enhancements (LLIE) of MSI and HSI followed by their fusion, a deep Unfolding HSI Super-Resolution with Automatic Exposure Correction (UHSR-AEC) is proposed, that can effectively generate a high-quality fused HSI-SR (in texture and features) even under very imbalanced exposures, thanks to the correlation between LLIE and HSI-SR taken into account.  Extensive experiments are provided to demonstrate the state-of-the-art overall performance of the proposed  UHSR-AEC, including comparison with some benchmark peer methods.

\end{abstract}
\section{Introduction}
\label{sec:intro}

Hyperspectral image (HSI) can capture the continuous spectral detection signal at individual pixel locations of an object \cite{green1998imaging, xie2016hyperspectral}. With rich information from hundreds of spectral bands, it has been widely used in remote sensing for target recognition \cite{ren2003automatic}, geological surveying \cite{ting2012application}, etc.
Owing to limited solar irradiance and hardware, there exists a compromise between the spatial and spectral resolution of HSI \cite{dian2017hyperspectral}. In contrast, normally multispectral image (MSI) has higher spatial resolution and lower acquisition difficulty than HSI. Fusion of MSI and HSI has been regarded as an effective HSI super-resolution (HSI-SR) approach \cite{wei2016multiband}. 

Hyperspectral sensors need to capture high quality HSI throughout the day. However, under extreme conditions such as night or poorly illuminating scenarios, the captured HSI and MSI are usually affected by low visibility, spectral distortion, etc., consequently degrading the spatial and spectral information quality and thus compromising the subsequent applications \cite{li2022low}. At the same time, due to the different sensing methods used by hyperspectral cameras and multispectral cameras, it is difficult to maintain the same exposure level for both HSI and MSI \cite{xie2020mhf, liu2009study}, thereby seriously affecting the quality of useful information in HSI and MSI, such that existing HSI-SR methods cannot perform effectively. One possible approach is to perform low-light image enhancement (LLIE) on both HSI and MSI, separately, and then carry out HSI-SR. However, this approach treats LLIE and HSI-SR
as two independent problems without considering their mutual effects in nature. This motivates our study for a new HSI-SR model that can mine mutual correlation, priors, and causal effects between LLIE and HSI-SR.

Existing HSI-SR methods can be categorized into two groups: model-based methods and deep learning-based methods. Model-based methods are based on the linear observation model linking the observed image and the original image, which relies on manually setting priors canonical rules. Yokoya et el. \cite{yokoya2011coupled} proposed a coupled nonnegative matrix factorization to estimate the abundance matrix and endmember of the HSI; Dong et al. \cite{dong2016hyperspectral} converted the estimation of HSI into a joint estimation of dictionary and sparse coding through the spectral sparsity of HSI; Dian et el. \cite{dian2019hyperspectral} proposed
a novel subspace-based low tensor multi-rank regularization method for HSI-SR. While model-based methods have explanations based on theoretical foundations, they often have poor performance due to their limited priors in well-defined mathematical forms. Deep learning methods have been widely used in HSI-SR in recent years. Xie et el. \cite{xie2020mhf} proposed an unfolding-based network by combining the low-rank prior and generalized model of HSI; Zhang et el. \cite{zhang2020deep} proposed a blind fusion network, which can overcome the mismatch between spectral and spatial responses. However, most existing HSI-SR methods have not yet seriously considered the degradation of complex environments, and the conditions for application are relatively limited. How to design new models to make HSI-SR applicable in complex environments is still an unsolved problem.

\begin{figure}[!htb]
\centering
\includegraphics[width=1\linewidth]{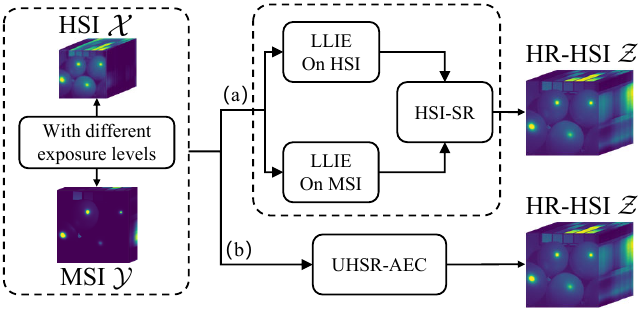}
\caption{Block diagram of generating HR-HSI $\cal Z$ from the acquired HSI $\cal X$ and MSI $\cal Y$ under different exposure levels, for (a) most existing methods and (b) the proposed  UHSR-AEC.}
\label{net0}
\end{figure}

Low-light image enhancement methods on natural images can be categorized into two groups: model-based methods and deep learning-based methods. Model-based LLIE methods focus on the statistical characteristics of the data. Ibrahim et al. \cite{ibrahim2007brightness} proposed a brightness-preserving dynamic histogram equalization method; Jobson et el. \cite{jobson1997properties} used the Retinex theory in the field of image brightness enhancement. However, hand-picked constraints often limit reconstruction performance, especially in challenging light-degraded scenes. Meanwhile, many deep learning-based methods have been widely applied to the LLIE field in recent years. Chen et el. \cite{wei2018deep} proposed a deep network structure based on Retinex theory; Wu et el. \cite{wu2022uretinex} combined deep unfolding and Retinex theory to good effect; Wang et el. \cite{wang2023exposurediffusion} proposed a diffusion-based network structure and a denoising process based on a physical exposure model, which can accurately capture noise and exposure. Deep learning-based LLIE methods are gradually becoming mainstream and have shown amazing performance. Since image enhancement methods on natural images do not take into account the spectral correlation of HSI, proposing an image enhancement method suitable for HSI remains an open challenge.

To solve the above problems, we propose a new HSI-SR restoration model, which integrates LLIE and SR, and can solve the HSI fusion in different exposure scenarios, as shown in Fig. \ref{net0}. At the same time, we design a deep unfolding method, which integrates the advantages of the model-based method and the deep learning-based method, and can solve the above problems. 

The main contributions are summarized as follows:
\begin{enumerate}
    \item By integrating the LLIE and SR problems of HSI, a new HSI-SR degradation and recovery model is proposed, which is essential to solving the problem of different exposures in HSI fusion. Then a novel deep Unfolding HSI Super-Resolution method with Automatic Exposure Correction (UHSR-AEC) based on deep unfolding is proposed to solve the problem.

    \item The proposed UHSR-AEC trains the considered model by solving three sub-problems (each with a data-fitting error and a regularizer) that are solved by the proximal gradient descent (PGD) algorithm, together with an Initialization Module (IM) for preserving details and texture features in HSI-SR.
    
    \item Extensive experiments are performed to demonstrate the effectiveness of the proposed  UHSR-AEC, including its state-of-the-art HSI-SR fusion performance and comparison with some existing benchmark LLIE-SR based methods. 
\end{enumerate}

\section{NOTATIONS AND PROBLEM FORMULATION}
\subsection{Notations}
In this paper, a scalar, a vector, a matrix, and a tensor are denoted by lowercase $x$, boldface lowercase $\mathbf{x}$, boldface capital $\mathbf{X}$, and calligraphic $\mathcal{X}$, respectively. $\mathbf{C}=\mathbf{A} \circ \mathbf{B}$ denotes the matrix-formed element-wise multiplication of $\mathbf{A}$ and $\mathbf{B}$, where matrices $\mathbf{A}$, $\mathbf{B}$, and $\mathbf{C}$ have the same dimension. $\|\mathbf{A}\|_\text{F}$ and $\|\mathbf{A}\|_1$ denote Frobenius norm and $\ell_1$ norm of matrix $\mathbf{A}$, respectively. For a tensor $\mathcal{X} \in \mathbb{R}^{I_1 \times I_2 \times \cdots \times I_N}$, the mode-$i$ unfolding martix is defined as $\mathbf{X}_{(i)} \in \mathbb{R}^{I_i\times \prod_{j\neq i}I_j}$.

\begin{figure*}[!htb]
\centering
\includegraphics[width=1\linewidth]{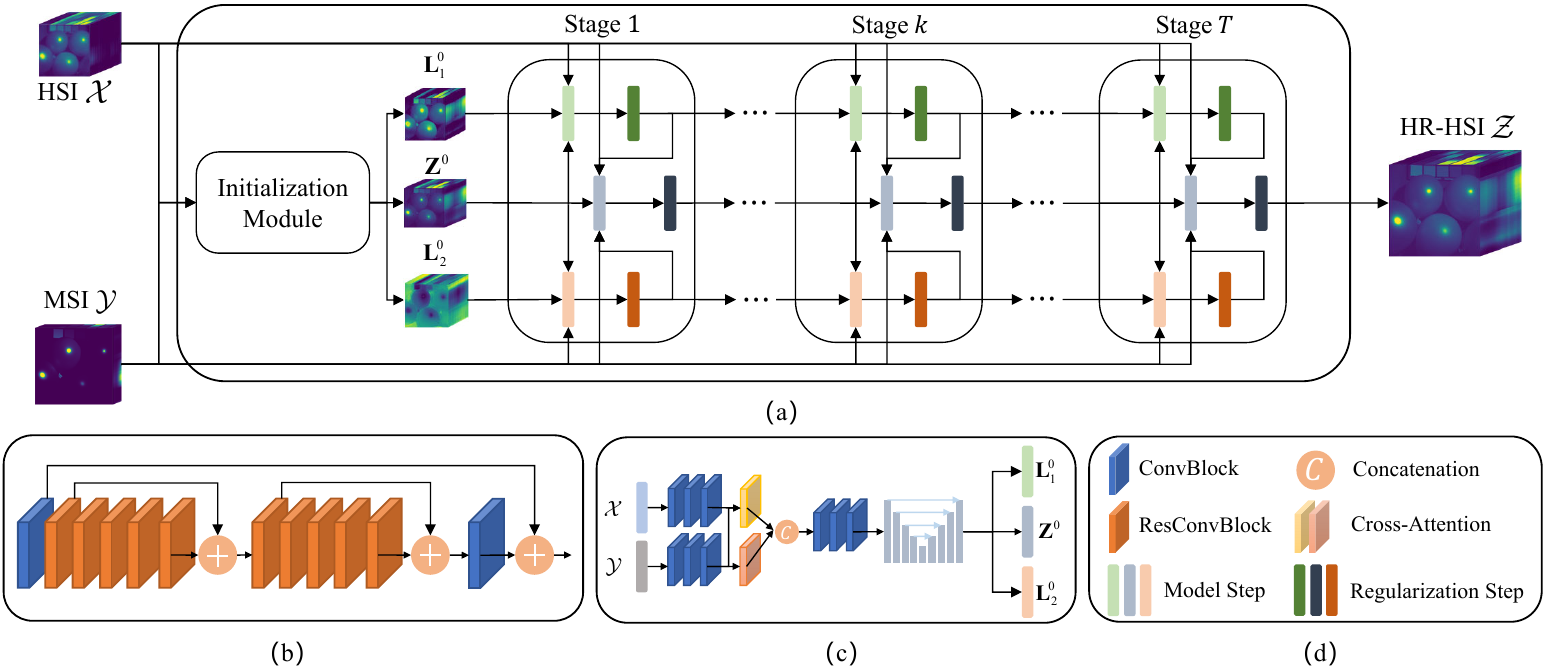}
\vspace{-2em}
\caption{(a) The proposed  UHSR-AEC, which consists of Initialization Module and Unfolding Module; (b) regularization step module, comprising two identical ResConvBlocks; (c) Initialization Module, comprising ConvBlock, Cross-attention, and Unet; (d) legend in the proposed  UHSR-AEC structure.}
\label{net}
\end{figure*}

\subsection{HSI-SR Degradation Model}
HSI-SR aims to recover an HR-HSI $\mathcal{Z} \in \mathbb{R}^{C\times W\times H}$ from an LR-HSI $\mathcal{X} \in \mathbb{R}^{C\times W_\text{HSI}\times H_\text{HSI}}$ and an HR-MSI $\mathcal{Y} \in \mathbb{R}^{C_\text{MSI}\times W\times H}$, where $C$ and $C_\text{MSI}$ are the spectral numbers in HSI and MSI, respectively. The degradation model for LR-HSI and HR-MSI can be expressed as follows:

\begin{equation}
\begin{aligned}
&\mathbf{X}_{(1)} = \mathbf{Z}_{(1)}\mathbf{H}\\
&\mathbf{Y}_{(1)} = \mathbf{P}\mathbf{Z}_{(1)}
\end{aligned}
\end{equation}
where $\mathbf{X}_{(1)} \in \mathbb{R}^{C\times N_{\text{HSI}}}$, $\mathbf{Y}_{(1)} \in \mathbb{R}^{C_{\text{MSI}}\times N}$, $\mathbf{Z}_{(1)} \in \mathbb{R}^{C\times N}$ are obtained by a mode-1 unfolding of $\mathcal{X}$, $\mathcal{Y}$, $\mathcal{Z}$, respectively, $N = W\times H$, $N_\text{HSI} = W_\text{HSI}\times H_\text{HSI}$ represent the number of pixels in $\mathbf{X}_{(1)}$ and $\mathbf{Y}_{(1)}$, respectively, $\mathbf{H} \in \mathbb{R}^{N \times N_\text{HSI}}$ is a spatial degeneracy matrix, which accounts for blurring and spatial downsampling operations, and $\mathbf{P} \in \mathbb{R}^{C_\text{MSI} \times C}$ is the spectral response matrix associated with the imaging sensor. For notational simplicity, in the following we will denote $\mathbf{X}_{(1)}, \mathbf{Y}_{(1)}, \mathbf{Z}_{(1)}$ by $\mathbf{X}, \mathbf{Y}, \mathbf{Z}$.\par

\section{PROPOSED METHOD}



\subsection{Exposure Correction HSI Fusion Model}

Considering the different exposure levels of the observational data $\mathcal{X}$ and $\mathcal{Y}$, we design a new HSI fusion model, which can be expressed as:

\begin{equation}
\begin{aligned}
\mathop{\min}\limits_{\mathbf{L}_1, \mathbf{L}_2, \mathbf{Z}}&\frac{1}{2}\|\mathbf{X} -(\mathbf{Z} \circ \mathbf{L}_1)\mathbf{H}\|_\text{F}^2 + \frac{1}{2}\|\mathbf{Y} - \mathbf{P}(\mathbf{Z} \circ \mathbf{L}_2)\|_\text{F}^2 \\
&+ \beta_1 \Phi_1(\mathbf{L}_1)+ \beta_2 \Phi_2(\mathbf{L}_2)+ \beta_3 \Phi_3(\mathbf{Z})
\end{aligned}
\label{model1}
\end{equation}
where $\mathbf{L}_1, \mathbf{L}_2 \in \mathbb{R}^{C\times N}$ represent the different levels of exposure on $\mathbf{Z}$, respectively, and $\Phi_1(\cdot)$, $\Phi_2(\cdot)$, $\Phi_3(\cdot)$ represent implicit regularizers for $\mathbf{L}_1$, $\mathbf{L}_2$, $\mathbf{Z}$, respectively. Note that $\mathbf{Z} \circ \mathbf{L}_1$ and $\mathbf{Z} \circ \mathbf{L}_2$ represent the light degraded $\mathbf{Z}$ at two different exposure settings, respectively. 

An iterative method for  solving problem (\ref{model1}) is proposed by solving three subproblems as follows:

\begin{equation}
\begin{aligned}
&\mathbf{L}_1=\mathop{\arg\min}\limits_{\mathbf{L}_1}\frac{1}{2}\|\mathbf{X} -(\mathbf{Z} \circ \mathbf{L}_1)\mathbf{H}\|_\text{F}^2+ \beta_1 \Phi_1(\mathbf{L}_1)\\
&\mathbf{L}_2=\mathop{\arg\min}\limits_{\mathbf{L}_2}\frac{1}{2}\|\mathbf{Y} - \mathbf{P}(\mathbf{Z} \circ \mathbf{L}_2)\|_\text{F}^2 + \beta_2\Phi_2(\mathbf{L}_2)\\
&\begin{aligned}\mathbf{Z}=\mathop{\arg\min}\limits_{\mathbf{Z}}&\frac{1}{2}\|\mathbf{X} -(\mathbf{Z} \circ \mathbf{L}_1)\mathbf{H}\|_\text{F}^2 \\
+ &\frac{1}{2}\|\mathbf{Y} - \mathbf{P}(\mathbf{Z} \circ \mathbf{L}_2)\|_\text{F}^2 + \beta_3 \Phi_3(\mathbf{Z})
\end{aligned}
\end{aligned}
\label{model2}
\end{equation}

At the iteration $t+1$, applying the PGD algorithm to the three subproblems in (\ref{model2}) yields:


\begin{equation}
\begin{aligned}
&\mathbf{L}_1^{t+1} = \text{Prox}_{\beta_1\Phi_1}(\mathbf{L}_1^t+\beta_1\mathbf{Z}^{t} \circ ((\mathbf{X}-(\mathbf{Z}^{t}\circ \mathbf{L}_1^t)\mathbf{H})\mathbf{H}^T))\\
&\mathbf{L}_2^{t+1} = \text{Prox}_{\beta_2\Phi_2}(\mathbf{L}_2^t+\beta_2\mathbf{Z}^{t} \circ (\mathbf{P}^T(\mathbf{Y}-\mathbf{P}(\mathbf{Z}^{t}\circ \mathbf{L}_2^t))))\\
&\begin{aligned}\mathbf{Z}^{t+1}=\text{Prox}_{\beta_3\Phi_3}(\mathbf{Z}^t &+ \beta_3((\mathbf{X}-(\mathbf{Z}^t\circ \mathbf{L}_1^{t+1})\mathbf{H})\mathbf{H}^T\circ \mathbf{L}_1^{t+1} \\
+& \mathbf{P}^T(\mathbf{Y}-\mathbf{P}(\mathbf{Z}^t\circ \mathbf{L}_2^{t+1}))\circ \mathbf{L}_2^{t+1}))
\end{aligned}\\
\end{aligned}
\label{model4}
\end{equation}
where $t=1,2,\cdots, T-1$, and $T$ is the maximum number of iterations, and $\text{Prox}_{\beta\Phi}(\cdot)$ denotes the proximal operator defined as
\begin{equation}
\text{Prox}_{\beta\Phi}(\mathbf{A}) = \mathop{\arg\min}\limits_{\mathbf{M}}\frac{1}{2}\|\mathbf{A}-\mathbf{M}\|_F^2+\beta\Phi(\mathbf{M})
\end{equation}

\subsection{Deep Unrolling on  UHSR-AEC}
The iterative updating stage in (\ref{model4}) can be implemented by the proposed  UHSR-AEC learning network as shown in Figure \ref{net}(a), through $T$ identical unfolding stages for regularizers' learning after the initialization. Next, let us present the structure of the proposed  UHSR-AEC in more detail, followed by the initialization module.


\textbf{Network Structure for Implicit Regularizer:} 
Figure \ref{net} (b) shows the designed residual recovery network (RRN) for implicit regularizers. Three RRNs with identical structure for three respective regularizers $\Phi_1(\cdot), \Phi_2(\cdot), \Phi_3(\cdot)$ are used at each learning stage. Each RRN consists of two identical ResConvBlocks, where the residual structure is considered to avoid the gradient vanishing and the performance improvement.


\textbf{Network Structure for Sampling Matrix:} To overcome the mismatch between the spatial and spectral responses used in the training and test data, we design learning sampling matrices to replace $\mathbf{H}, \mathbf{H}^T, \mathbf{P}, \mathbf{P}^T$, where $\mathbf{P}$ and $\mathbf{P}^T$ ($\mathbf{H}$ and $\mathbf{H}^T$) each contains one Conv (three
ConvBlocks).

\subsection{Initialization Module}
Randomly initialized $\mathbf{L}_1^0$, $\mathbf{L}_2^0$, $\mathbf{Z}^0$ may destroy the texture and details of the image, resulting in degradation of the recovery quality. Therefore, the initialization module (IM) is designed for better initial $\mathbf{L}_1^0$, $\mathbf{L}_2^0$, $\mathbf{Z}^0$.

Figure \ref{net}(c) shows the network structure of IM, composed of a feature extraction (FE) layer, a cross-attention (CA) layer, and a feature fusion (FF) layer. The FE first projects $\mathbf{X}$ and $\mathbf{Y}$ onto the high-dimensional space, and upsamples $\mathbf{X}$ and $\mathbf{Y}$ in the spatial domain and the spectral domain, respectively; the CA extracts the Channel Attention of $\mathbf{X}$ and the Spatial Attention of $\mathbf{Y}$ for high-quality spectral and spatial information; the FF layer formed by  a Unet, fuses features of $\mathbf{X}$ and $\mathbf{Y}$ and maps them to $\mathbf{L}_1^0$, $\mathbf{L}_2^0$, $\mathbf{Z}^0$, by solving the following problem:

\begin{equation}
\begin{aligned}
\mathop{\min}\limits_{\mathbf{L}_1^0, \mathbf{L}_2^0, \mathbf{Z}^0}&\|\hat{\mathbf{Z}}-\mathbf{Z}^0\|_1 + \lambda\|\hat{\mathbf{Z}}_x-(\mathbf{Z}^0 \circ \mathbf{L}_1^0)\|_1\\
+&\lambda\|\hat{\mathbf{Z}}_y-(\mathbf{Z}^0 \circ \mathbf{L}_2^0)\|_1
\end{aligned}
\label{loss0}
\end{equation}
where $\hat{\mathbf{Z}}$, $\hat{\mathbf{Z}}_x$ and $\hat{\mathbf{Z}}_y$ represent the ground truth, the associated exposure degraded HSI and MSI to mitigate the degradation of exposure, respectively. Note that IM is important in the training of UHSR-AEC, but it is virtual as the trained UHSR-AEC operates for the reconstruction of HSI-SR. 


\subsection{Loss Function} 
The proposed  UHSR-AEC is trained through an end-to-end approach, while IM is fine-tuned at first during the training, together with the weights shared among $\Phi_1(\cdot)$, $\Phi_2(\cdot)$, and $\Phi_3(\cdot)$ at different sequential unfolding stages. The optimization problem with a $\ell_1$-norm based loss function, considered for training  UHSR-AEC by its greater robustness against outliers than the Frobenius norm, is as follows:  
\begin{equation}
\begin{aligned}
\mathop{\min}\limits_{\mathbf{L}_1^T, \mathbf{L}_2^T, \mathbf{Z}^T} &\|\hat{\mathbf{Z}} - \mathbf{Z}^T\|_1 + \eta_1\|\hat{\mathbf{Z}}_x - (\mathbf{Z}^T \circ \mathbf{L}_1^T)\|_1\\
+&\eta_1\|\hat{\mathbf{Z}}_y - (\mathbf{Z}^T \circ \mathbf{L}_2^T)\|_1 + \eta_2\|\hat{\mathbf{X}} - (\mathbf{Z}^T \circ \mathbf{L}_1^T)\mathbf{H}\|_1 \\
+&\eta_2\|\hat{\mathbf{Y}} - \mathbf{P}(\mathbf{Z}^T \circ \mathbf{L}_2^T)\|_1
\end{aligned}
\label{loss1}
\end{equation}
\vspace{-2em}

\begin{table*}[]
\centering
\caption{Quantitative results of various methods in terms of PSNR and SSIM (SAM and ERGAS) for which ${\uparrow}$” (${\downarrow}$”) indicates that the larger (smaller) the numerical values, the better the corresponding results. The best results are shown in boldface and the second-best results are underlined.}
\label{Table2}
\resizebox{\linewidth}{!}{%
\begin{tabular}{c|cccc|cccc|cccc|cccc}
\hline
\multirow{2}{*}{Methods} & \multicolumn{4}{c|}{CAVE (Case 1)} & \multicolumn{4}{c|}{CAVE (Case 2)} & \multicolumn{4}{c|}{Harvard (Case 1)} & \multicolumn{4}{c}{Harvard (Case 2)} \\
                         & PSNR$\uparrow$    & SSIM$\uparrow$   & SAM$\downarrow$   & ERGAS$\downarrow$   & PSNR    & SSIM   & SAM   & ERGAS   & PSNR    & SSIM    & SAM    & ERGAS    & PSNR    & SSIM    & SAM    & ERGAS   \\ \hline
LIME+LTMR                &11.7235       &0.4291      &\underline{11.0445}     &67.1560       &14.1318       &0.5953      &20.1203     &52.2342       &7.1238       &0.1832       &\underline{8.9612}      &163.9742        &9.1335       &0.2298       &16.6631      &142.3775       \\
LIME+LTTR                &11.7232       &0.4300      &11.1318     &67.1966       &14.1554       &0.6014      &20.0475     &52.2843       &7.0917       &0.1820       &9.1761      &165.6484        &8.0263       &0.1959       &17.2600      &162.9281       \\
LIME+MoG-DCN             &11.8697       &0.4152      &12.6203     &65.2987       &14.2794       &0.5722      &19.8989     &51.1158       &5.5263       &0.1039       &26.8368      &369.3135        &5.8667       &0.1362       &22.5598      &258.6686       \\
RetinexNet+LTMR          &22.9439       &0.7653      &14.2370     &16.9788       &21.3002       &0.7027      &16.7437     &19.6997       &23.0712       &0.6509       &9.6458      &25.0398        &19.6607       &0.5548       &14.4514      &86.1875       \\
RetinexNet+LTTR          &22.9460       &0.7655      &14.2807     &16.9592       &21.2952       &0.7036      &16.8666     &19.7264       &23.1197       &\underline{0.6549}       &9.6127      &25.1389        &19.6356       &0.5564       &\underline{14.2904}      &86.5815       \\
RetinexNet+MoG-DCN       &23.0481       &\underline{0.7725}      &13.6159     &\underline{16.3326}       &21.3092       &\underline{0.7063}      &15.8519     &19.4943       &23.2560       &0.6515       &10.8736      &24.0247        &19.3001       &0.5057       &15.8946      &\underline{83.0595}       \\
EFINet+LTMR              &\underline{23.0812}       &0.7158      &12.7621     &16.8142       &\underline{21.8557}       &0.6918      &13.1353     &\underline{18.3252}       &22.9766       &0.5929       &32.3514      &24.1489        &19.3429       &0.5681       &16.3868      &121.7584       \\
EFINet+LTTR              &23.0637       &0.7163      &12.9124     &16.8624       &21.8375       &0.6920      &13.2972     &18.3738       &\underline{23.4273}       &0.6219       &31.8069      &22.6756        &19.3349       &0.5679       &16.4290      &122.1559       \\
EFINet+MoG-DCN           &22.9020       &0.7193      &12.7179     &17.1430       &21.7530       &0.6949      &\underline{13.0449}     &18.5050       &23.2625       &0.6345       &23.9204      &\underline{21.1815}        &\underline{19.7600}       &\underline{0.5759}       &16.2332      &112.1969       \\
 UHSR-AEC                     & \textbf{27.1015}    & \textbf{0.8601}    & \textbf{9.4557}   & \textbf{11.0631}       & \textbf{25.0026}       & \textbf{0.8355}      & \textbf{10.1323}     & \textbf{12.8321}       & \textbf{32.6981}       & \textbf{0.9227}       & \textbf{6.1609}      & \textbf{8.1784}        & \textbf{30.0148}       & \textbf{0.8885}       & \textbf{6.6361}      & \textbf{10.3762}       \\ \hline
\end{tabular}%
}
\end{table*}

\begin{figure*}[!htb]
\centering
\includegraphics[width=1\linewidth]{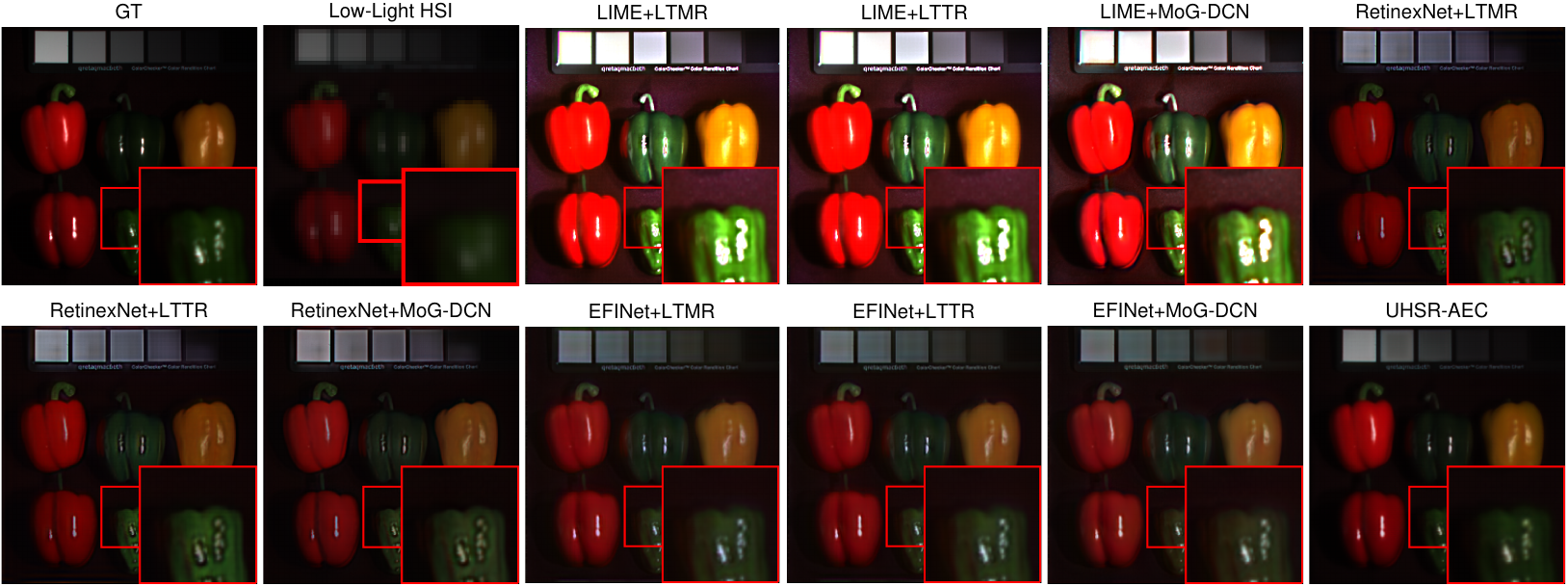}
\vspace{-2em}
\caption{Reconstructed images of various methods for the fake\_and\_real\_peppers\_ms (CAVE) dataset are illustrated by the false color image of [30,15,10] bands in Case 1.}
\vspace{-0.5em}
\label{cave}
\end{figure*}

\begin{figure*}[!htb]
\centering
\includegraphics[width=1\linewidth]{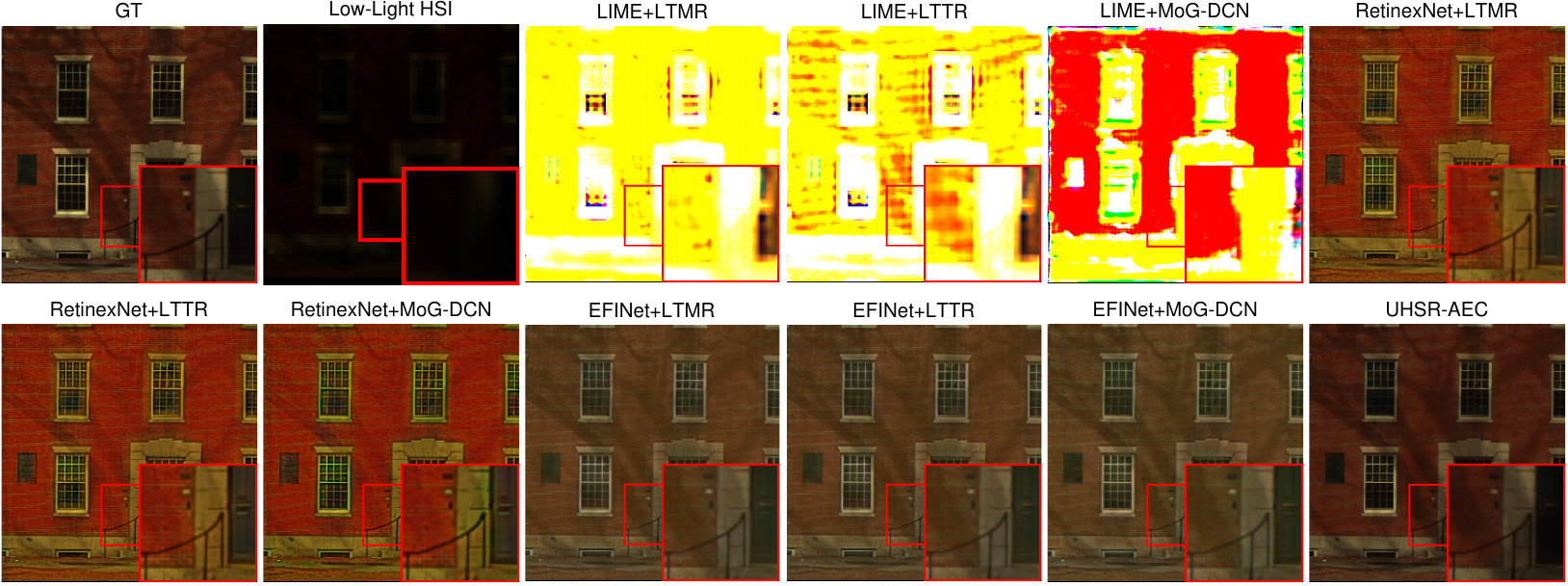}
\vspace{-2em}
\caption{Reconstructed images of various methods for the Imge3 (Harvard) dataset are illustrated by the false color image of  [30,15,10] bands in Case 2.}
\vspace{-0.5em}
\label{harvard}
\end{figure*}

\section{Experiments}
\subsection{Experimental settings}
\textbf{Datasets:} The datasets CAVE\footnote[1]{https://www1.cs.columbia.edu/CAVE/databases/multispectral/} and Harvard\footnote[2]{http://vision.seas.harvard.edu/hyperspec/explore.html} are used to evaluate the effectiveness of the proposed method. The CAVE dataset contains 32 HSIs, each with 31 spectral bands and 512$\times$512 pixels, and we select 20 HSIs as the training dataset and 12 HSIs as the test dataset; the Harvard dataset contains 50 HSIs, each with 31 spectral bands and 1392$\times$1040 pixels, and we select 30 HSIs as the training dataset and 20 HSIs as the test dataset. Due to the limited computational resources, we crop and downsample each training HSI into 64$\times$64 pixels and each testing HSI into 256$\times$256 pixels.\\
\textbf{Comparison methods and evaluation indicators:} 
Peer methods compared with the proposed  UHSR-AEC include three LLIE methods: LIME \cite{LIME}, RetinexNet \cite{wei2018deep}, and EFINet \cite{EFINet}; and three HSI-SR methods: LTMR \cite{dian2019hyperspectral}, LTTR \cite{LTTR}, and MoG-DCN \cite{dong2021model}. For consistency, the deep learning-based methods are all trained with the same HSI datasets.

In the experiments, all the HSI data are normalized to [0,1]. Four metrics are chosen to evaluate reconstructed HSI-SR image quality: peak signal-to-noise ratio (PSNR), structural similarity index (SSIM), spectral angle mapper (SAM), and the dimensionless global relative error of synthesis (ERGAS). All the experiments are performed on a server with a GPU NVIDIA 4090 graphics card.\\
\textbf{Implementation Details:} For each reference image, i.e., HR-HSI $\mathbf{Z}$ (GT), we adjust the image exposure by two different values of Gamma correction ($\alpha \in U(0.2,2)$, $\gamma \in U(0.5,3)$), for generating $\hat{\mathbf{Z}}_x$ and $\hat{\mathbf{Z}}_y$; the observation $\mathbf{X}$ is generated from $\hat {\mathbf{Z}}_x$ via Gaussian blurring (Gaussian kernel size of 8, mean of 0, standard deviation of $\sqrt{3}$) and downsampling (downsampling ratio of $K = |\mathbf{Z}|/|\mathbf{X}| = 4$); the observation $\mathbf{Y}$ is generated through the convolutional integration of $\hat{\mathbf{Z}}_y$  with a real simulated spectral response (based on a Nikon camera).

We use the Pytorch framework to implement the proposed method. In the proposed  UHSR-AEC network, we use a four-layer Unet structure and the Adam optimizer with the learning rate $l_r = 10^{-4}$, as well as the maximum number of iterations $I_{max} = 250,000$. Moreover, $\beta_1, \beta_2, \beta_3, \lambda, \eta_1, \eta_2, T$ are set to 0.001, 0.001, 0.005, 0.5, 0.3, 0.1, 3, respectively. We would like to mention that the loss functions in (\ref{loss0}) and (\ref{loss1}) are for one HSI training sample $(\hat{\bf Z},\hat{\bf Z}_x,\hat{\bf Z}_y)$, while for the case of multiple HSI training samples in our experiment, they are simply replaced by the sum of the corresponding loss functions associated with each sample \cite{dong2021model}.

\subsection{Comparison with competitive methods}
The simulation HSI and MSI data are generated using CAVE and Harvard datasets for the following two exposure degradation cases:

Case 1: ($\alpha_1$ = 0.5,$\gamma_1$ = 0.7) for HSI, and ($\alpha_2$ = 1.3,$\gamma_2$ = 1.5) for MSI. 

Case 2: ($\alpha_1$ = 0.5,$\gamma_1$ = 2.0) for HSI, and ($\alpha_2$ = 0.8,$\gamma_2$ = 1.5) for MSI. 

The obtained simulation results are shown in Table \ref{Table2}. As can be seen from this table,  UHSR-AEC is overall superior to the other nine methods. For the visual quality assessment, Figs. \ref{cave} and \ref{harvard} show some results for all tested methods on the CAVE and Harvard datasets, respectively. For each sub-image, the reconstructed HR-HSI for three spectral bands ([30,15,10]) is shown. From these sub-images, it can be observed that: i) LIME-based methods yield overly bright images, ii) RetinexNet and EFINet-based methods expose some detail loss and colour imbalance in Case 1 and Case 2, respectively, iii)  UHSR-AEC shows the best visualization in both Case 1 and Case 2.
\vspace{-1em}

\subsection{Ablation studies}
Some results for Ablation experiments on the proposed  UHSR-AEC are shown in Table \ref{ablation}, which are obtained using the CAVE dataset with exposure degradation parameters used in Case 1. One can see from this table, that the proposed  UHSR-AEC performs worst without IM, while it (with IM) performs best only in terms of PSNR for $T=3$, and best for $T=4$ otherwise, indicating that IM is essential and a suitable value of $T$ is also needed for the proposed  UHSR-AEC to operate in good shape.

\vspace{-1em}

\section{Conclusion}
We have presented a novel UHSR-AEC network for HSI-SR (as shown in Fig. \ref{net}), with the given HSI and MSI acquired under different exposures. The proposed  UHSR-AEC, 
with the structure for deep unfolding for intelligent image fusion and IM for preserving image texture and details,  
is devised from a new degradation-restoration perspective, instead of two independent domains (i.e., LLIE and restoration) considered by most existing approaches. Extensive experiments are provided to demonstrate the superior overall performance of the proposed UHSR-AEC over some existing LLIE-SR based benchmark methods. 

\begin{table}[]
\centering
\caption{ UHSR-AEC ablation experimental results for performance sensitivity to IM module and system parameter $T$.}
\label{ablation}
\resizebox{\columnwidth}{!}{%
\begin{tabular}{ccccc}
\hline
Methods     & PSNR$\uparrow$ & SSIM$\uparrow$ & SAM$\downarrow$ & ERGAS$\downarrow$ \\ \hline
 UHSR-AEC ($T$=3)   & \textbf{27.1015}    & \underline{0.8601}    & \underline{9.4557}   & \underline{11.0631}     \\ \hline
 UHSR-AEC ($T$=2)   & 26.5207    & 0.8484    & 10.1317   & 11.7043    \\
 UHSR-AEC ($T$=4)   & \underline{26.9205}    & \textbf{0.8827}    & \textbf{9.0812}   & \textbf{10.7911}     \\ \hline
 UHSR-AEC w/o IM ($T$=3) & 22.6863    & 0.7259    & 12.4081   & 17.2980     \\ \hline
\end{tabular}%
}
\end{table}

\bibliographystyle{IEEEbib}
\bibliography{icme2023template}

\end{document}